\documentstyle[12pt,a4,epsfig]{article}
  \def\thebibliography#1{\center{\bf REFERENCES}\list
   {[\arabic{enumi}]}{\settowidth\labelwidth{[#1]}\leftmargin\labelwidth
   \advance\leftmargin\labelsep
   \usecounter{enumi}}
   \def\newblock{\hskip .11em plus .33em minus -.07em}
   \sloppy
   \sfcode`\.=1000\relax}
  
  \newcounter{cap}
  {\begin{list}{Figure \arabic{cap}\hfil}{\usecounter{cap}
  \settowidth{\labelwidth}{Figure #1}%
  \setlength{\leftmargin}{\labelwidth}%
  \addtolength{\leftmargin}{\labelsep}%
  \setlength{\parsep}{2mm plus 1mm minus 1mm}
  \setlength{\itemsep}{3mm plus 2mm minus 2mm}
  }}%
  {\end{list}}
  {\begin{list}{}{\settowidth{\labelwidth}{#1}%
  \setlength{\leftmargin}{\labelwidth}%
  \addtolength{\leftmargin}{\labelsep}%
  \setlength{\itemsep}{0pt plus 1pt}
  \setlength{\parsep}{0pt plus 1pt}
  \setlength{\topsep}{0pt plus 1pt}
  \setlength{\partopsep}{0pt plus 1pt}
  \setlength{\parskip}{2mm plus 1mm minus 1mm}
  }}%
  {\end{list}}
\def\sd{\strut\displaystyle}

\newcommand{\be}{\begin{equation}}
\newcommand{\ee}{\end{equation}}
\newcommand{\ggppp}{$\gamma \gamma \to \pi \pi \pi$}

\begin{document}
\parindent 1.3cm
\begin{titlepage}

\begin{flushright}
NORDITA 95/79 N,P\\
UAB-FT-95/381\\
UG-FT-56/95\\
hep-ph/9512296
\end{flushright}

\begin{center}
\begin{bf}
\noindent
$\gamma \gamma \to \pi \pi \pi$ TO ONE LOOP
IN  CHIRAL PERTURBATION THEORY 
\end{bf}
  \vspace{1.5cm}\\
P. TALAVERA and Ll. AMETLLER
\vspace{0.1cm}\\
Departament F{\'\i}sica i Enginyeria Nuclear, UPC,\\
08034  (Barcelona), Spain\\
\vspace{0.5cm}
J. BIJNENS
  \vspace{0.1cm}\\
NORDITA, Blegdamsvej 17 \\
DK-2100 Copenhagen  \O, Denmark\\
\vspace{0.5cm}
A. BRAMON
   \vspace{0.1cm}\\
Grup de F\'\i sica Te\`orica, Universitat Aut\`onoma de Barcelona,\\
08193 Bellaterra (Barcelona), Spain\\
   \vspace{0.5cm}
and \\
\vspace{0.5cm}
F. CORNET
\vspace{0.1cm}\\
Departamento de F\'\i sica Te\'orica y del Cosmos,\\
Universidad de Granada, 18071 Granada, Spain\\
   \vspace{2.4cm}

{\bf ABSTRACT}
\end{center}

The $\gamma\gamma \to \pi^0 \pi^0 \pi^0$ and
$\gamma\gamma \to \pi^+ \pi^- \pi^0$ amplitudes are discussed
in the general context of Chiral Perturbation Theory (ChPT) to $O(p^6)$.
Chiral loops are found to play a major role. This makes these processes
a good test of ChPT, mainly in its anomalous sector.
We correct earlier numerical results at tree
level and determine the one-loop results as well.

\end{titlepage}

\section{INTRODUCTION}
The purpose of the present paper is to discuss the $\gamma \gamma \to
\pi^0 \pi^0 \pi^0$ and  $\gamma \gamma \to
\pi^+ \pi^- \pi^0$ transitions at the one loop level in Chiral
Perturbation Theory (ChPT) \cite{GASSER}. The experimental interest on
these transitions centers in projected high-luminosity
$e^+e^-$-machines (such as the DAPHNE $\phi$-factory under
construction in Frascati) from which data can be expected in a near
future. In the past, these two processes were treated in the
theoretical context of Current Algebra (CA). Adler et al \cite{Adler}
solved a preexisting controversy on the $\gamma \gamma \to \pi \pi \pi$
amplitudes and, more recently, their CA results
for the  $\gamma \gamma \to \pi^+ \pi^- \pi^0$ amplitude have been
confirmed by Bos et al \cite{Bos} working with ChPT at tree
level, {\it i.e}., in a context which is essentially equivalent to CA.
The situation, however, is not satisfactory.
First, because there are good reasons to believe that
one loop corrections to the lowest order amplitudes (which vanish in
a specific chiral limit \cite{Adler}) can be very important. Secondly,
because the predictions in Ref. \cite{Bos} for the  $\gamma \gamma \to
\pi^+ \pi^- \pi^0$ {\it cross section} are in sharp disagreement with (roughly,
one order of magnitude smaller than) those coming from previous work
using equivalent lowest order amplitudes \cite{Pratap}.

In the context of 2 flavor ChPT,
the triplet of pseudoscalar mesons $P$, {\it i.e.}, the pions, is described in
terms
of the Hermitian  matrix
\begin{equation}
  \label{M_8}
P=\pmatrix{
{\sd\pi^0 \over \sd\sqrt 2} &  \pi^+   \cr
\pi^-   & -{\sd\pi^0 \over \sd\sqrt 2} \cr
 },
\end{equation}
which appears in the various pieces of the ChPT lagrangian through
the conventional para\-me\-tri\-za\-tion $
\Sigma\equiv \exp\left(\sd{2i P/f}\right) $,
with $f$ at lowest order equal to
the charged pion decay constant $f
=f_{\pi}=132~\hbox{MeV}$
\cite{GASSER,PDG}.

The lowest order lagrangian of ChPT (order two in particle four--momenta
or masses, $O(p^2)$ ) is
\begin{equation}
\label{L2}
{\cal L}_2=\sd{f^2\over 8} tr (D_\mu\Sigma D^\mu\Sigma^\dagger + \chi
\Sigma^\dagger +\chi^\dagger \Sigma).
\end{equation}
The covariant derivative,
$D_\mu \Sigma \equiv \partial_\mu \Sigma + i e A_\mu [Q,\Sigma]$,
contains the photon field $A_\mu$ and the quark charge matrix $Q$
[$Q=\hbox{diag}(2/3,-1/3)$], and the non--derivative terms
are proportional to the quark mass matrix via the identification
$\chi=\chi^\dagger = B~\cal M$, with [${\cal M}=\hbox{diag}(m_u,m_d)$].
The parameter $B$ relates at lowest order the pion mass with the quark
masses: $B = 2{m_\pi^2}/(m_u+m_d)$.

The next order lagrangian, $O(p^4)$, can be divided in two sectors:
the Wess--Zumino term \cite{WESS} and a series of
seven terms identified and studied by
Gasser and Leutwyler \cite{GASSER},
\begin{equation}
\label{L4}
{\cal L}_4= {\cal L}_{WZ} + \sum_{i=1}^{7} {l}_i {\cal L}_4^{(i)}.
\end{equation}
The only pieces of ${\cal L}_{WZ}$ relevant for the
$\gamma \gamma \to \pi \pi \pi$ amplitudes are
\begin{eqnarray}
\label{WZ}
{\cal L}_{WZ}=&-& \sd{e\over 16\pi^2}\epsilon^{\mu\nu\alpha\beta} A_\mu
tr ( Q
\partial_\nu\Sigma\partial_\alpha\Sigma^\dagger
\partial_\beta\Sigma \Sigma^\dagger-
Q\partial_\nu\Sigma^\dagger\partial_\alpha\Sigma\partial_
\beta\Sigma^\dagger \Sigma) \\
&-&i\sd{e^2\over 8\pi^2}\epsilon^{\mu\nu\alpha\beta} \partial_\mu
A_\nu A_\alpha tr (Q^2 \partial_\beta \Sigma \Sigma^\dagger
+Q^2  \Sigma^\dagger \partial_\beta \Sigma - {1\over 2}Q \Sigma Q
 \partial_\beta  \Sigma^\dagger + {1\over 2} Q
 \Sigma^\dagger Q  \partial_\beta  \Sigma ).
\nonumber
\end{eqnarray}
The seven non-anomalous terms of ${\cal L}_4$ contain products of covariant
derivatives and/or mass--terms similar to those in Eq.(\ref{L2}). Each one of
the corresponding constants $l_i$ can be divided into two pieces: a divergent
one and a finite, real constant. The divergent terms are needed to cancel
the divergences appearing in one--loop calculations with vertices from
${\cal L}_2$, thus rendering the results finite,
while the renormalized real constants,
$l_i^r$, have been fixed by experimental data \cite{GASSER}.
Alternatively, the various $l_i^r$
can be deduced with a good approximation assuming that they are
saturated by the exchange of known meson resonances
as discussed in Refs. \cite{ECKER,DONOGHUE}. Fixing the renormalization
mass--scale around these resonance masses ($\mu=M_\rho$,
for instance), the finite and renormalized values for $l_i^r$ are
small enough to justify the convergence of
the perturbative series. These remarks obviously do not apply
to the ${\cal L}_{WZ}$ term in Eq. (\ref{L4}). It generates only
anomalous processes with coupling strengths derived from the
anomaly. However,
higher order (counter-) terms belonging to ${\cal L}_6$ -- which
can be similarly separated into two pieces contributing to anomalous
and non-anomalous processes, respectively --
are in general expected to give smaller contributions
than their corresponding lower order lagrangians
in Eq. (\ref{L4}): ${\cal L}_{WZ}$ and
$l_i^r{\cal L}_4^{(i)}$.
Details on counterterms in the anomalous sector and the
saturation of the corresponding free constants in terms of
resonances can be found in Refs.
\cite{BIJNENS1,BIJNENS2,HANS,AKHOURY}. 

\section{TREE-LEVEL AMPLITUDE FOR \ggppp}

The lowest order contribution in ChPT (order four in particle four--momenta
or masses)  to the amplitude for
$\gamma \gamma \to \pi^0 \pi^0 \pi^0$ proceeds exclusively through the
first diagram of Fig.1, where the propagator corresponds to a $\pi^0$.
One can immediately read the two relevant
vertices from the lagrangians ${\cal L}_2$ and ${\cal L}_{WZ}$
in Eqs. (\ref{L2})
and (\ref{WZ}). If $p_{1,2,3}$ denote the three pion four--momenta and
$k, k'$ and ${\epsilon, \epsilon '}$ the two photon four--momenta and
polarizations, one easily obtains
\be
\label{A000}
A(\gamma \gamma \to \pi^0 \pi^0 \pi^0)^{tree} =
{e^2 \over \sqrt 2 \pi^2 f_\pi^3} {m_\pi^2 \over s-m_\pi^2} \epsilon_{\mu \nu
\alpha \beta}  \epsilon^\mu k^\nu \epsilon'^\alpha k'^\beta ,
\ee
where $s=(k+k')^2=(p_1+p_2+p_3)^2$ and $m_\pi^2=p_i^2$. Notice that this
amplitude is proportional to $m_\pi^2$, thus vanishing in the chiral limit.
The corresponding cross section for $\gamma \gamma \to \pi^0 \pi^0 \pi^0$
at this (lowest) order has been plotted (dashed line) in Fig.2.
\par
The lowest order amplitude for $\gamma \gamma \to \pi^+ \pi^- \pi^0$
receives contributions from all the diagrams in Fig. 1 and
can similarly be obtained to be
\begin{eqnarray}
\label{A+-0}
A(\gamma \gamma \to \pi^+ \pi^- \pi^0)^{tree} &=&
{e^2 \over \sqrt 2 \pi^2 f_\pi^3} \epsilon_{\mu \nu \alpha \beta}
\epsilon^\mu k^\nu \Bigl[{1 \over 2}\bigl(-1 + {p_{+-}^2 - m_\pi^ 2
\over s-m_\pi^2} \bigr) \epsilon'^\alpha k'^\beta
\nonumber\\
 &+&\bigl(
\epsilon'^\alpha -{p_+\cdot \epsilon' \over p_+\cdot k' } p_-^\alpha
 -{p_- \cdot \epsilon' \over p_-\cdot  k' } p_+^\alpha \bigr) p_0^\beta \Bigr]
+ \Bigl[{k\choose \epsilon} \leftrightarrow {k'\choose \epsilon'}
\Bigr]  ,
\end{eqnarray}
where now the three pion four--momenta are written as $p_+, p_- $
and $p_0$ and the notation $p_{ij}^2\equiv (p_i+p_j)^ 2$ is used.
Our tree-level amplitude coincides with those previously deduced
in Refs. \cite{Adler} and \cite{Bos}.
The corresponding cross section has been plotted in Fig.3 (dashed line).
It turns out to be one order of magnitude smaller than the cross
section predicted by Bos et al. \cite{Bos} and two orders of magnitude
smaller than that in Ref. \cite{Pratap}. 
The $3\pi^0$ and $\pi^+\pi^-\pi^0$  cross-sections are similar
near threshold but the one for the neutral pions is much smaller for
large values of the center of mass energy. This is due to the $m_\pi^2$
proportionality of the amplitude in Eq. (\ref{A000}).
\par
In order to check our results and to understand the origin of the
discrepancies with \cite{Bos} and \cite{Pratap},
we have performed an analytic calculation of the
tree-level cross sections for $\gamma \gamma \to \pi^0 \pi^0 \pi^0$
and $\gamma \gamma \to \pi^+ \pi^- \pi^0$ in the non--relativistic
approximation (NR). We take $m_{\pi^{\pm}}=m_{\pi^0}$ for simplicity and
for illustration purposes.
For both cross sections one has
\be
\label{NR}
\sigma (\gamma \gamma \to \pi \pi \pi) ^{NR} =
{1 \over 2^7 3 \sqrt3 \pi^2} \left( 1-6{m_{\pi} \over \sqrt s}
+9{m^2_{\pi} \over s}
\right) |T|^2 ,
\ee
where the squared matrix elements are
\be
\label{000}
|T_{000}|^2 = {1 \over 3!} {1 \over 4} \Sigma
|A (\gamma \gamma \to \pi^0 \pi^0 \pi^0)|^2 = {1 \over 3!}({\alpha
m^2_\pi \over \pi f^3})^2 ({s \over s - m^2_\pi})^2 \to
{1 \over 3!}({9 \alpha m^2_\pi \over 8 \pi f^3})^2 ,
\ee
for the neutral pion case, and
\be
\label{+-0}
|T_{+-0}|^2 =  {1 \over 4} \Sigma
|A (\gamma \gamma \to \pi^+ \pi^- \pi^0)|^2 \to
 18 ({\alpha m^2_\pi \over \pi f^3})^2 \left[ \left( {7 \over4} + {1
\over 128} \right) + 2 - \left( 4 -{1 \over 4} \right) \right] =
({3 \alpha m^2_\pi \over 8 \pi f^3})^2 ,
\ee
for the charged one. In both cases, the arrow ($ \to $) indicates that
we have restricted to values at threshold, $s = 9m_\pi^2$. This allows for
several tests. From Eq. (\ref{NR}) and the $s$ dependent values in
Eq. (\ref{000}) one obtains a reasonable approximation to the $3 \pi^0$
cross section at the tree-level for the whole range of relevant
energies, as shown (dot-dashed line) in Fig.2. Similarly, from Eq. (\ref{NR})
and the threshold value in Eq. (\ref{+-0}) one obtains the short solid line
drawn near threshold  in Fig.3, in good agreement with our tree-level
prediction for this charged pion case. Notice also that the threshold
value for the $\gamma \gamma \to \pi^0 \pi^0 \pi^0$ amplitude is found
to be three times larger than that for $\gamma \gamma \to
\pi^+ \pi^- \pi^0$ in agreement with Ref. \cite{Adler}.
\par
A plausible explanation for the huge disagreement between our
predictions for $\sigma(\gamma \gamma \to
\pi^+ \pi^- \pi^0)$ and those from previous work (Refs. \cite{Pratap}
and \cite{Bos}) is offered by Eq. (\ref{+-0}). There is a drastic
destructive interference between the two different amplitudes shown in
Eq. (\ref{A+-0}) as seen inside the brackets in Eq. (\ref{+-0}), where the
first two terms refer to the two squared moduli and the third, negative
one, to the interference. To see that this drastic effect is also
valid for the whole range of energies, the contribution to the
cross-section from each
independent amplitude has been plotted in Fig.3 (two almost
coincident higher lines) compared to the total cross-section.
A precise and numerically accurate treatment
is required to extract the correct values for the tree-level cross
section from the difference between these two large and destructive
contributions.

\section{$O(p^6)$ CORRECTIONS}

The one-loop contributions in ChPT (order six in particle four--momenta
or masses)  to the amplitudes for
$\gamma \gamma \to \pi \pi \pi$ proceed through various
(not--shown) diagrams and from wave function, mass and decay constant
renormalization. We have neglected the effects of the $\eta$ and
kaon loops, since they are expected to be small as in
$\gamma\gamma\to\pi\pi$\cite{BIJCOR}.


We have performed the calculation of the diagrams contributing to the
$\gamma \gamma \to \pi^0 \pi^0 \pi^0$ amplitude
both by hand and
using the algebraic manipulation program FORM \cite{FORM}.
One obtains the same result which turns out to be divergent
and requires appropriate counterterms.
However, inspecting the $O(p^6)$ counterterms to
anomalous processes given in Ref. \cite{BIJNENS1} one realizes that there
is none contributing to this process. Thus, the divergences
(generated only by loops with non-anomalous vertices) must be
cancelled by the $l_i$ counterterms in Eq. (\ref{L4}).
A convenient expression is given in terms of the
scale independent $\bar l$'s, introduced by Gasser and Leutwyler
\cite{GASSER} in $SU(2)_L\times SU(2)_R$ ChPT. The matrix element at
$O(p^6)$ reads
\begin{eqnarray}
\label{p000oneloop}
& & A(\gamma \gamma \to \pi^0 \pi^0 \pi^0) =
{e^2 \over 2\sqrt 2 \pi^2 f_\pi^3} {m_\pi^2 \over s-m_\pi^2} \epsilon_{\mu \nu
\alpha \beta} \nonumber \\
&\Biggl\{& \Biggl[\epsilon^\mu k^\nu \epsilon'^\alpha k'^\beta
 \Biggl( {1 \over 3} + {1 \over 16 \pi^2 f_\pi^2}
\Bigl[ (2\bar{l}_1 + 4\bar{l}_2 -6) ({3p_{12}^4 -
s^2 -3m_\pi^4 \over 9m_\pi^2} )
\nonumber\\&&
 -{2 \over 3}(\bar{l}_4-1)(s-3m_\pi^2)
-(\bar{l}_3-1)m_\pi^2
\nonumber \\
& &  +\Bigl( 
2(s-p^2_{12})({p^2_{12}\over m_\pi^2}-1)-m_\pi^2\Bigr) N(p^2_{12})
-8(s-m_\pi^2)({p^2_{12}\over m_\pi^2}-1)
R(p^2_{12},k'\cdot p_{12}) \Bigr] \Biggr) + \nonumber \\
&+&  \epsilon^\mu k^\nu {1 \over 16 \pi^2 f_\pi^2}
\Biggl( 8(s-m_\pi^2) \bigl( {p^2_{12} \over m_\pi^2} - 1 \bigr)
R(p^2_{12}, k'\cdot  p_{12}) \bigl( \epsilon'^\alpha -
{ \epsilon' \cdot p_{12} \over k' \cdot p_{12}} k'^\alpha \bigr)  p_{12}^\beta
\Biggr) \nonumber \\
&&+ \Biggl(p_{12} \leftrightarrow p_{13}
\Biggr)  + \Biggl(p_{12} \leftrightarrow p_{23} \Biggr) \Biggr]
+ \Biggl[{k\choose \epsilon} \leftrightarrow {k'\choose \epsilon'}
 \Biggr] \Biggr\} ,
\end{eqnarray}
where
\begin{eqnarray}
N(p^2) & \equiv & - \beta \ln { \beta - 1 \over \beta + 1} - 2
\nonumber \\
R(p^2, k\cdot p) & \equiv & {I(\lambda^2) - I(\lambda^{\prime 2})
\over \lambda^2 - \lambda^{\prime 2}} + {1 \over 2}N(p^2)
\end{eqnarray}
with
\begin{eqnarray}
\lambda^2 & \equiv & p^2/m_\pi^2, \ \ \lambda^{\prime 2} \equiv (p^2 -
2p\cdot k)/m_\pi^2, \ \ \beta \equiv \sqrt {1 - 4/ \lambda^2} \nonumber \\
I(\lambda^2) & \equiv & {1 \over 2} \ln^2 {\beta - 1 \over \beta + 1}
+ {3 \over 2} \lambda^2 +  {\beta \lambda^2 \over 2}
\ln {\beta - 1 \over \beta + 1} ,
\end{eqnarray}
Notice that there are three independent gauge invariant amplitudes.
The first one is of the tree-type (\ref{A000})
and it is the only one that requires the
introduction of the counterterms, reflected in the presence of the finite
constants $\bar{l}_i$.
There is also no contribution from the VMD estimate
of the ``anomalous'' $O(p^6)$ counterterms.
Moreover, it should be stressed that,
contrary to the tree level amplitude, the $O(p^6)$ result is no longer
proportional to $m_\pi^2$. There is thus no reason to expect small
corrections to the lowest order cross-section
when including the $O(p^6)$ contribution.


The number of one-loop diagrams contributing to the
$\gamma \gamma \to \pi^+ \pi^- \pi^0$ amplitude is larger than in the
neutral channel. Thus, we have performed this longer calculation using
FORM \cite{FORM}. Some partial checks, however, have been done using partial
subsets of diagrams. We obtained the
$\pi \pi$ scattering amplitudes as given by Gasser and Leutwyler \cite{GASSER}
and the $\gamma \pi\pi\pi$ amplitude derived in \cite{BIJNENS2}. The loops
give a gauge invariant result, with divergent contributions which have
to be cancelled by appropriate counterterms.
Contrary to the neutral process, where
the non-anomalous $O(p^4)$ $ l_i$ were the only counterterms needed,
the charged process needs additional, genuine, $O(p^6)$ counterterms.
These have been discussed in general in
Ref. \cite{BIJNENS1,AKHOURY,SCHERER}. We have
explicitly checked that the divergences
appearing in the one-loop calculation cancel with the known counterterms from
the previous references. The terms of $O(p^6)$ from the
lagrangian also contribute
to the amplitude via tree diagrams that contain two free constants.
These constants have been fixed assuming their saturation by the vector
meson contribution \cite {BIJNENS1}.

The final expression for the amplitude is very long and will be given
elsewhere \cite{WE}. We will only mention here that it contains $10$
independent amplitudes (actually, using Schouten identities it can be
shown that this is the maximum number of allowed independent amplitudes),
whereas only $3$ of them appeared at lowest order.

\section{NUMERICAL RESULTS}

The $O(p^6)$ cross-sections for  $\gamma \gamma \to \pi^0 \pi^0 \pi^0$
and  $\gamma \gamma \to \pi^+ \pi^- \pi^0$ 
have been plotted (solid lines) in Figs. 2 and 3 respectively, 
We have used $m_{\pi^\pm} = m_{\pi^0}$ in the amplitude but the experimental
values in the phase space and we fixed the renormalization scale $\mu=M_\rho$,
according to our assumption of the saturation of the free constants
in the lagrangian at $O(p^6)$ by
the vector meson resonances. The values used for the constants in
${\cal L}_4$ that contribute to our processes are the central values quoted in
\cite{GASSER},
\begin{equation}
\bar l_1=-2.3 \quad \bar l_2=6.0 \quad \bar l_3=2.9 \quad \bar l_4=4.3 .
\end{equation}

The corrections are very large in both channels. In the neutral channel the
corrections
increase the cross-section up to two orders of magnitude! This is due
to the vanishing of the lowest order amplitude in the chiral limit,
which no longer occurs at $O(p^6)$. Since the $O(p^4)$
amplitude is proportional to $m_\pi^2$, which is very small
compared to the momenta involved in the process, the corrections are very 
large.
In the charged channel the reason for the small cross-section at lowest
order is different. Here there is a large cancellation between the two
gauge invariant amplitudes contributing to this process. The $O(p^6)$
corrections modify both amplitudes, thus spoiling the almost perfect
cancellation, and adds new gauge invariant amplitudes.

The $\gamma \gamma \to \pi\pi\pi$
cross sections obtained at one loop in ChPT are significantly larger
than the lowest order predictions. However,
they are still smaller than the ones for other
interesting $\gamma \gamma$ processes as, for instance,
$\gamma \gamma \to \pi^0 \pi^0$ \cite{BIJCOR,DHL,BGS}.
In any case, they have some
chances of being measured at Daphne. We have
estimated that, working with the optimal projected machine luminosity,
about $180$ $\pi^+ \pi^- \pi^0$
and $23$ $\pi^0 \pi^0 \pi^0$ events per year should originate from
photon--photon collisions. An eventual experimental confirmation of these
processes would be a clear indication of the important role played by the
chiral loops in the anomalous WZ sector.

In summary, we have estimated the cross sections for the processes
$\gamma \gamma \to \pi^+\pi^-\pi^0$ and $\gamma \gamma \to \pi^0\pi^0\pi^0$
at lowest order, $O(p^4)$, and at one loop, $O(p^6)$, in Chiral Perturbation
Theory.
The corrections are extremely large due to the smallness of the lowest
order cross-sections. Since the reasons for these small values at lowest
order are understood and disappear at $O(p^6)$ in both channels,
it is not expected that
the $O(p^8)$ corrections will modify our results in an important way.

\par
ACKNOWLEDGEMENTS

We are pleased to thank D. Espriu, Ll. Garrido, J. Gasser, J. Matias
and J. Taron for discussions. This work was partially supported by
CICYT under contracts: AEN94-0936 and AEN95-0815, and by EURODAPHNE,
HCMP, EEC Contract \#CHRX-CT920026.


\listoffigures
\newpage

\begin{figure}
\input FEYNMAN
\vspace{1cm}

\begin{picture}(10000,32000)(0,-16000)


  \drawline\photon[\NE\REG](0,6000)[8]
  \drawline\photon[\SE\REG](-160,16000)[8]
  \drawline\scalar[\E\REG](\photonbackx,\photonbacky)[6]
  \drawline\scalar[\SE\REG](11000,11100)[3]
  \drawline\scalar[\NE\REG](11000,11100)[3]

  \put (8000,15000){(a) \hspace{7cm} (b)}


  \drawline\photon[\NE\REG](25000,6000)[8]
  \drawline\photon[\SE\REG](25000,16000)[8]
  \drawline\scalar[\E\REG](\photonbackx,\photonbacky)[4]
  \drawline\scalar[\SE\REG](\photonbackx,\photonbacky)[4]
  \drawline\scalar[\NE\REG](\photonbackx,\photonbacky)[4]

  \put (8000,600){(c)  \hspace{7cm} (d)}


  \drawline\photon[\NE\REG](0,-10000)[8]
  \drawline\scalar[\E\REG](\photonbackx,\photonbacky)[4]
  \drawline\scalar[\SE\REG](\photonbackx,\photonbacky)[4]
  \drawline\scalar[\NE\REG](\photonbackx,\photonbacky)[4]
  \drawline\photon[\SE\REG](2500,+2500)[8]
  \drawline\photon[\NE\REG](25000,-10000)[8]
  \drawline\scalar[\E\REG](\photonbackx,\photonbacky)[4]
  \drawline\scalar[\SE\REG](\photonbackx,\photonbacky)[4]
  \drawline\scalar[\NE\REG](\photonbackx,\photonbacky)[4]
  \drawline\photon[\NE\REG](28000,-13000)[8]

  \end{picture}
\caption{Tree level diagrams in ChPT for $\gamma \gamma \to
\pi\pi\pi$.}
\end{figure}
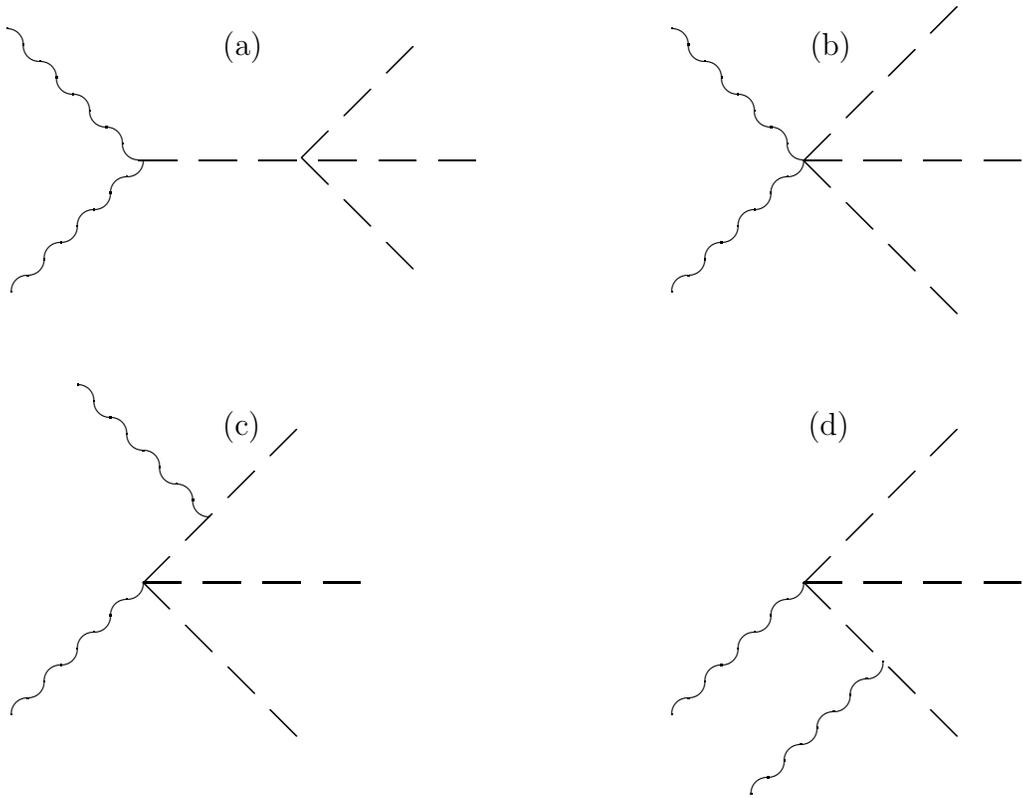
\begin{figure}
\epsfig{file=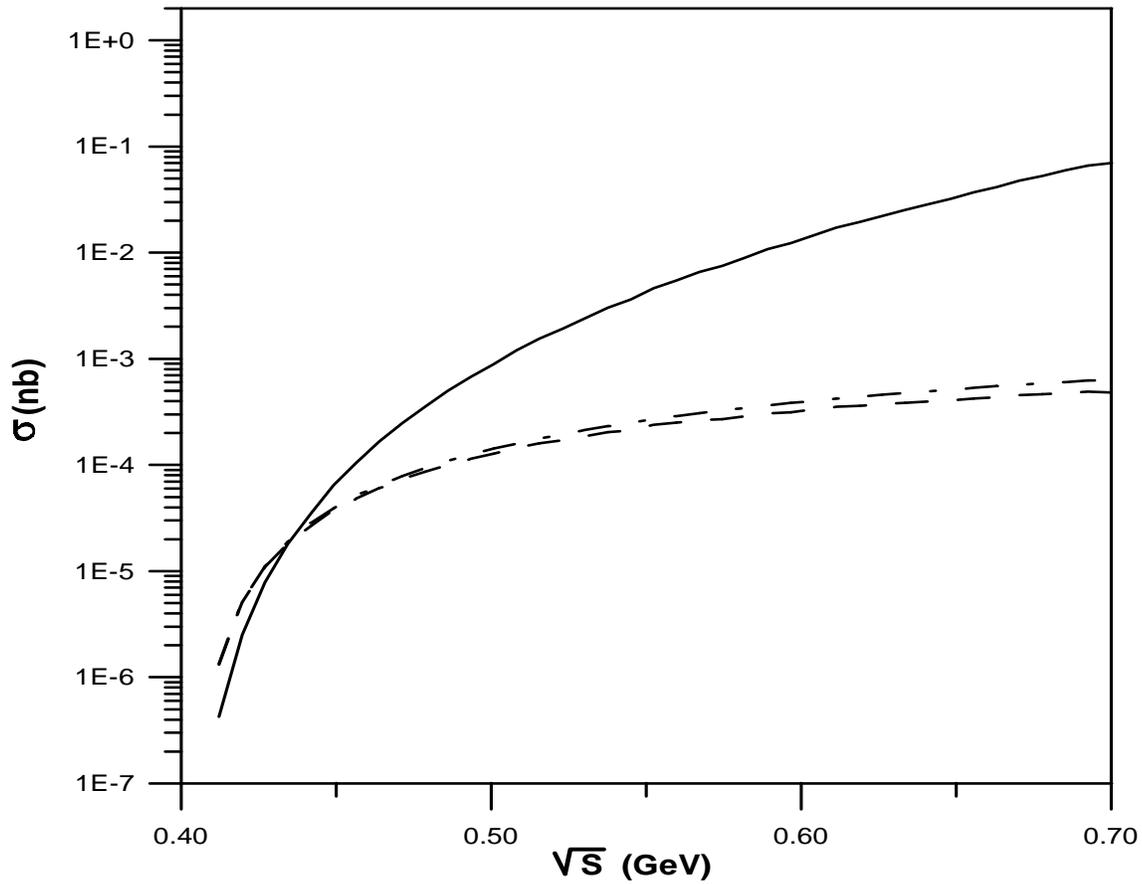,width=16cm,height=19cm,angle=0}
\caption[$\gamma \gamma \to
\pi^0\pi^0\pi^0$ cross section.]{$\gamma \gamma \to
\pi^0\pi^0\pi^0$ cross section at tree level (dashed line)
as a function of $\sqrt{s}$. The dot-dashed line
corresponds to the non--relativistic tree level approximation.
The solid line corresponds to the
$O(p^6)$ result.}
\end{figure}
\begin{figure}
\epsfig{file=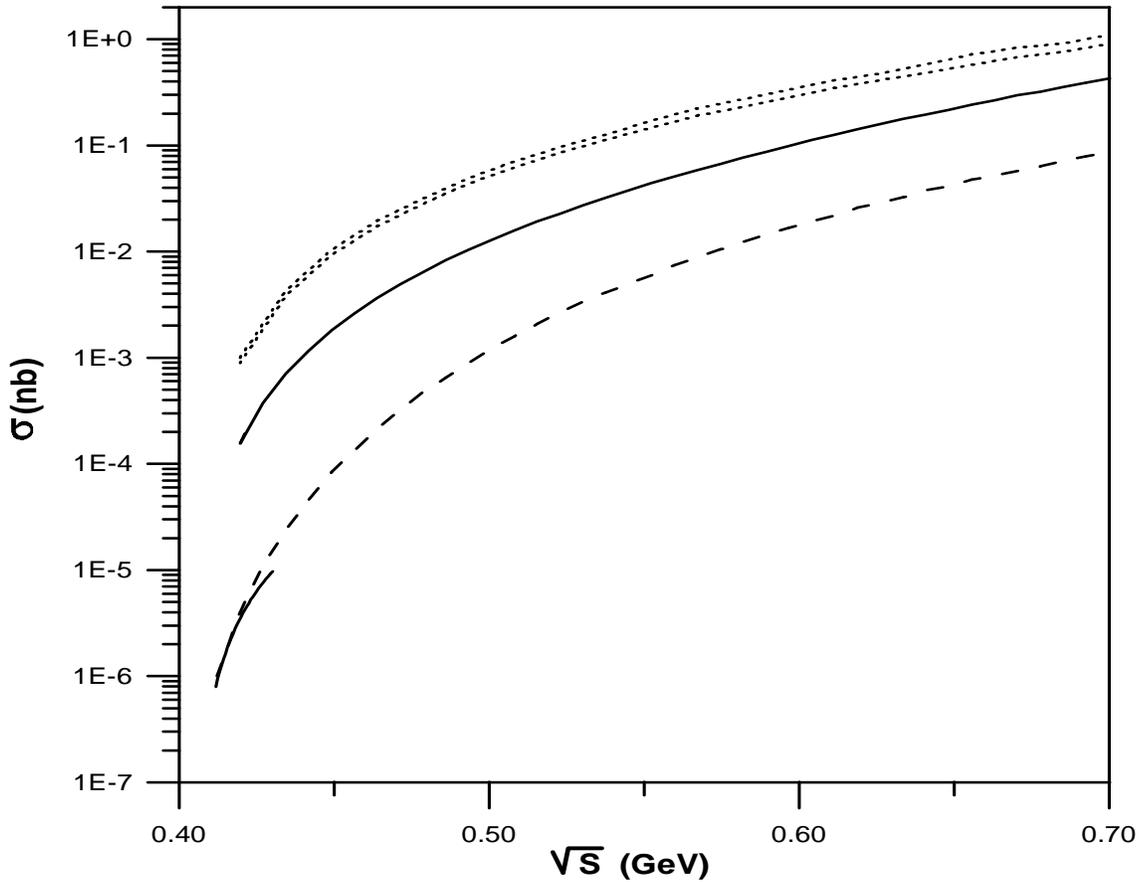,width=16cm,height=19cm}
\caption[
$\gamma \gamma \to \pi^+\pi^-\pi^0$ cross section]{$\gamma \gamma \to \pi^+\pi^-\pi^0$ cross section
as a function of
$\sqrt{s}$. The dashed line corresponds to the tree
level ChPT prediction. The short solid line near threshold is the
non--relativistic tree level approximation.
The two higher dotted curves show how would the cross section be
for each one of the two gauge invariant amplitudes in 
Eq.(\protect{\ref{A+-0}}),
and
illustrate their (largely destructive) interference effects.
The solid line corresponds to the $O(p^6)$ result.}
\end{figure}
\end{document}